\makeatletter
\declare@file@substitution{revtex4-1.cls}{revtex4-2.cls}
\makeatother

\documentclass[twocolumn]{aastex631}
\usepackage{amsmath}
\usepackage{graphicx}

\shorttitle{E/B power asymmetry of dust polarization}

\begin{document}

\title{Obtaining strength of magnetic field from E and B modes of dust polarization}


\author{Jungyeon Cho}
\affiliation{Department of Astronomy and Space Science, Chungnam National University, Daejeon, Republic of Korea}
\email{jcho@cnu.ac.kr}





\begin{abstract}
We perform numerical simulations of supersonic magnetohydrodynamic (MHD) turbulence 
and calculate Fourier power spectra of E and B modes arising from dust polarization. 
We pay close attention to  the ratio of E-mode to B-mode spectra (a.k.a. E/B power asymmetry) on small spatial scales.
We find that the ratio depends on the strength of the mean magnetic field: the stronger the mean magnetic field is, the smaller the ratio is. 
More precisely speaking, the ratio scales with the Alfv\'en Mach number $M_A$, the root-mean-square velocity divided by the Alfv\'en speed of the mean magnetic field, when it lies in the range $1\lesssim M_A \lesssim 30$.
This result implies that we can use the E/B power asymmetry to constrain the strength of the mean magnetic field in supersonic and super-Alfv\'enic MHD turbulence.

\end{abstract}


\section{Introduction} \label{sec:intro}
Magnetic field plays important roles in many astrophysical environments. 
In the interstellar medium (ISM), it affects key processes in star formation, such as condensation of cores and launch of outflows (Mac Low \& Klessen 2004;
McKee \& Ostriker 2007; Crutcher 2012; Federrath \& Klessen 2012; Bally 2016; Hu, Lazarian \& Stanimirovic 2021).
Therefore obtaining the strength of magnetic field is of great importance in understanding star formation processes.

Various methods exist to measure the strength of magnetic fields in astrophysical fluids (see, for example, Crutcher 2012).
Among these the use of dust polarization is most popular (see, for example, Ward-Thompson et al. 2017; Pattle et al. 2022).
Since interstellar dust grains are aligned mainly with respect to local magnetic field directions and, as a result, thermal  emission from the grains is polarized in the direction perpendicular to the magnetic field (see, for example, Lazarian, Andersson \& Hoang 2015), 
we can study magnetic field from polarization observations. 

The Davis-Chandrasekhar-Fermi method (Davis 1951; Chandrasekhar \& Fermi 1953; hereinafter DCF method) is a simple and  powerful technique for estimating
the strength of the mean magnetic field projected on the plane of the sky $B_{0,sky}$.
If we observe dispersion of the polarization angle $\delta \phi$ arising from magnetically aligned dust grains and the dispersion of line-of-sight (LOS) velocity $\delta v_{LOS}$, 
we can obtain  $B_{0,sky}$ using the DCF method:
\begin{equation}
 B_{0,sky} 
    = Q \sqrt{ 4 \pi \bar{\rho} } \frac{ \delta v_{LOS} }{\delta \phi },   \label{eq:trad}
\end{equation}
where $Q$ is a correction factor, which is usually taken as $\sim$0.5 
(Ostriker, Stone \& Gammie 2001; Padoan et al. 2001; Heitsch et al. 2001),
 and $\bar{\rho}$ is average density.
Due to its simplicity, the DCF method is widely used in the community
 (see, for example, 
Gonatas et al. 1990; Lai et al. 2001; Di Francesco et al. 2001;
Crutcher et al. 2004; Girart et al. 2006; Curran \& Chrysostomou 2007; 
Heyer et al 2008; Mao et al. 2008; Tang et al. 2009; Sugitani et al. 2011; Ward-Thompson et al. 2017).
Further elaboration of the DCF method has been made by many researchers 
(Zweibel 1990; Myers \& Goodman 1991; Zweibel 1996; Ostriker et al. 2001; 
Heitsch et al. 2001; Padoan et al. 2001; Kudoh \& Basu 2003; Wiebe \& Watson 2004;
Falceta-Gon{\c c}alves et al. 2008; Hildebrand et al. 2009; Houde et al 2009; Cho \& Yoo 2016; Yoon \& Cho 2019; Skalidis \& Tassis 2021; Chen et al. 2022; see also Lazarian, Yuen \& Pogosyan 2022 and Cho 2019).

The DCF method has some limitations.
For example, the number of independent eddies along the LOS should not be large.  
If there are many independent eddies along the LOS, the DCF method tends to overestimate $B_{0,sky}$ (Cho \& Yoo, 2016).
Another limitation is that the angular dispersion should not be large. 
Ostriker et al. (2001) showed that the DCF method yields a good estimates
of $B_{0,sky}$ when $\delta \phi \lesssim 25^\circ$.
Suppose that there is just one independent turbulence eddy along the LOS.
In this case, we have
\begin{equation}
  \delta \phi \sim \frac{ b }{B_0},
\end{equation}
where $b$ and $B_0$ are strengths of  the fluctuating and the mean magnetic  fields, respectively.
If we assume  $b$ and
the root-mean-square (rms) velocity $v_{rms}$ are related by $ b/\sqrt{4 \pi \bar{\rho}}  \sim v_{rms}$ as in Alfv\'{e}nic turbulence, we have
\begin{equation}
  \delta \phi \sim \sqrt{ 4 \pi \bar{\rho} }\frac{v_{rms}}{B_0}=\frac{v_{rms}}{ B_0/\sqrt{ 4 \pi \bar{\rho} }}
       = \frac{v_{rms}}{ V_{A0} }=M_{A},
\end{equation}
where $V_{A0}$ is the Alfv\'{e}n speed of the mean magnetic field and $M_{A}$ is the Alfv\'{e}n Mach number.
Note that $M_{A}$ can be a measure of the strength of the mean magnetic field. 
The mean magnetic field is strong in sub-Alfv\'{e}nic turbulence (i.e., turbulence with $M_{A} \lesssim 1$) and weak in super--Alfv\'{e}nic turbulence (i.e., turbulence with $M_{A} \gtrsim 1$).
Therefore, the DCF method works for sub-Alfv\'{e}nic turbulence with 
\begin{equation}
    M_{A} \lesssim \tan 25^\circ \sim 0.5,
\end{equation}
if there are not many independent turbulence eddies along the LOS.

In this paper, we report numerical results that can be used to constrain the value of $B_0$ in  supersonic and super-Alfv\'{e}nic turbulence.
Our calculation is based on E (gradient-type) and B (curl-type) modes of dust polarization, which provide rotational invariant description of polarization patterns. 
The Planck observations (Planck Collaboration XXX 2016) of dust polarization at 353 GHz found E/B power asymmetry: 
 E-mode power is roughly twice the B-mode power.
The Planck observations have triggered theoretical investigation of the E/B power asymmetry in magnetohydrodynamic (MHD) turbulence
(Caldwell et al. 2017; Kandel et al. 2017, 2018; Kritsuk et al. 2018; Kim et al. 2019; Brandenburg et al. 2019). 
Caldwell et al. (2017) and Kandel et al. (2017, 2018) analytically studied  the E/B power asymmetry and others numerically studied the asymmetry in multi-phase ISM turbulence (Kritsuk et al. 2018), Galactic disk (Kim et al. 2019), and turbulence in the early universe and solar surface (Brandenburg et al. 2019). 

In this paper, we numerically study the relation between the E/B power asymmetry and the strength of the mean magnetic field $B_0$ in supersonic MHD turbulence.
In \S2, we describe our numerical method.
In \S3, we present results of our numerical simulations.
In \S4, we give discussion and summary.

\section{Numerical Methods} \label{sect:num}

\subsection{Equations}
\label{sect:code}
To obtain turbulence data, we solve the following compressible MHD equations 
in a periodic box of size $2\pi$
 (see Cho \& Lazarian 2002):
\begin{eqnarray}
{\partial \rho    }/{\partial t} + \nabla \cdot (\rho {\bf v}) =0,  \\
{\partial {\bf v} }/{\partial t} + {\bf v}\cdot \nabla {\bf v} 
   +  \rho^{-1}  \nabla(C_s^2\rho)\nonumber\\
   - (\nabla \times {\bf B})\times {\bf B}/4\pi \rho ={\bf f},  \\
{\partial {\bf B}}/{\partial t} -
     \nabla \times ({\bf v} \times{\bf B}) =0, 
\end{eqnarray}
with $\nabla$$\cdot$$\bf B$$=$0 and an isothermal equation of state $P$=$C^{2}_{s}\rho$, 
where  $C_{s}$ is the sound speed, $P$ is pressure, and $\rho$ is density. 
Here $\bf{v}$ is the velocity, $\bf{B}$ is the magnetic field, 
and $\bf{f}$ is the driving force. 
The magnetic field consists of the mean field  and a fluctuating field: $\bf{B}$=$\bf{B}_0$+$\bf{b}$.
We use $512^3$ grid points.
In our simulations, $\bar{\rho}=1$ and the rms velocity $v_{rms}$ is roughly 1.
Therefore, the sonic Mach number ($M_s = v_{rms}/C_s$) is $\sim 1/C_s$ and  
the Alfv\'en Mach number of the mean magnetic field ($M_{A}= v_{rms}/V_{A0}$) is   $\sim 1/V_{A0}=\sqrt{4 \pi \bar{\rho}}/B_0$.

\subsection{Forcing}
In this work, we drive turbulence in Fourier space and consider only solenoidal ($\nabla \cdot {\bf f}=0$) forcing. 
We use two different driving scales: $k_f = L/L_f  \sim$ 2.5 and 10, where $L$ is the box size (=$2\pi$) and $L_f$ is the driving scale.
When $k_f \sim 2.5$, we use 22 forcing components isotropically distributed in the range $2 \lesssim k \lesssim \sqrt{12}$, where $k$ is the wavenumber.
When $k_f \sim 10$, we use 100 forcing components isotropically distributed in the range $10/1.3\lesssim k \lesssim 13$.
    More detailed descriptions on forcing can be found in Yoo \& Cho (2014).

\subsection{Simulation groups}
We perform 3 groups of simulations:
\begin{itemize}
\item {\bf G1}: Simulations with $M_s \sim8$ and $k_f \sim 2,5$.
          In all simulations, $C_s=0.1$ and $v_{rms} \sim 0.8$.
          In these simulations, the number of eddies along the LOS is roughly 2.5.
\item {\bf G2}: Simulations with $M_s \sim 11$ and $k_f \sim 10$. In all simulations, $C_s=0.1$ and $v_{rms} \sim 1.1$.
          In these simulations, the number of eddies along the LOS is roughly 10.
          The purpose of these simulations is to see the effects of the number of eddies along the LOS.
\item {\bf G3}: Simulations with $M_s \sim 3.5$ and $k_f \sim 10$. In all simulations, $C_s=\sqrt{0.1}$ and $v_{rms} \sim 1.1$.
          In these simulations, the number of eddies along the LOS is roughly 10.
           The purpose of these simulations is to see the effects of the sonic Mach number $M_s$.
\end{itemize}
Table \ref{tbl:sim} summarizes the simulation parameters and results. 

\subsection{Calculation of E/B power asymmetry} \label{sect:method}
We obtain the E/B power ratio from the following procedures.
\begin{enumerate}

\item We obtain three-dimensional turbulence data from direct numerical simulations (see \S \ref{sect:code}).
\item Using the method in Fiege \& Pudritz (2000;  see also Heitsch et al. 2001), we calculate Stokes parameters Q and U 
             arising from magnetically aligned dust grains at a far-infrared/sub-millimeter wavelength. 
             Except \S \ref{sect:30}, we assume the LOS is perpendicular to the mean magnetic field ${\bf B}_0$.
\item We obtain E and B modes in Fourier space via
      \begin{equation}
           \tilde{R}(k_x,k_y) = (\hat{k}_x -i \hat{k}_y)^2  \tilde{P}( k_x, k_y ),
      \end{equation}
where $R=E +iB$, $P=Q+iU$,  $\hat{k}_x = k_x/k$, $\hat{k}_y = k_y/k$, and $k=(k_x^2 +k_y^2)^{1/2}$. 
Tildes denote Fourier transformation in the sky plane (i.e., $xy$-plane). 
See Brandenburg et al. (2019) and references therein for details. 

\item We calculate E/B power ratio for small $spatial$ scales from
      \begin{equation}
           [E/B]_{small} =  \frac{ \int^{k_{max}}_{k_{min}}E_E(k) dk }{ \int^{k_{max}}_{k_{min}}E_B(k) dk},   \label{eq:eb_asym}
       \end{equation}
       where
       \begin{equation}
          E_E(k)=\int_{k-0.5}^{k+0.5} |\tilde{E}( k_x, k_y)|^2 dk
       \end{equation}
       and $E_B(k)$ is defined similarly. The value of $k_{max}$ is determined by the numerical resolution.
       We choose $k_{min}=20$. Although our choice of $k_{min}$ is arbitrary, our results are not very sensitive to $k_{min}$.
       
\item We investigate how the E/B power ratio changes as $B_0$  (i.e., the strength of the mean magnetic field) changes.

\end{enumerate}
Although we present results mostly for the case that the LOS is perpendicular to the mean magnetic field,
we also present results for the case that the LOS makes an angle of 30 degrees with the mean magnetic field (see \S \ref{sect:30}).

\begin{deluxetable*}{lcccccccc}
\tabletypesize{\scriptsize}
\tablecaption{Parameters of Simulations}
\tablewidth{0pt}
\tablehead{ 
\colhead{Run\tablenotemark{a}} & 
\colhead{$M_{s}$\tablenotemark{b}} &
\colhead{$M_{A}$\tablenotemark{c}} & 
\colhead{$V_{A0}$\tablenotemark{d}     } & 
\colhead{$v_{rms}$} & 
\colhead{$L/L_{f}$\tablenotemark{e}} &
\colhead{resolutions} &
\colhead{$<[E/B]_{small}>$ \tablenotemark{f}} &  
\colhead{$<[E/B]_{small,30}>$ \tablenotemark{g}} 
 } 
\startdata
G1-B0.005        &  8.6   &  171  &  0.005   & 0.858  &           $\sim$2.5   & $512^3$    & 2.18    & 2.24  \\
G1-B0.01        &   7.7   &  76.7 &    0.01  &  0.767  &            $\sim$2.5   & $512^3$    & 2.29     & 2.26  \\
G1-B0.05        &   8.1   &  16.2    &  0.05  & 0.810  &            $\sim$2.5   & $512^3$    & 1.90     & 1.88  \\
G1-B0.1        &   7.9   &  7.86   &  0.1   &   0.786  &            $\sim$2.5   & $512^3$    & 1.67     & 1.66  \\
G1-B0.2        &  7.5  &  3.77   &   0.2  &  0.755  &           $\sim$2.5   & $512^3$    & 1.45    & 1.46 \\
G1-B0.5        &   7.6  &  1.51   &  0.5   &  0.756  &            $\sim$2.5   & $512^3$    & 1.18     & 1.30  \\
G1-B1        &   8.2   &  0.815   &  1   &  0.815  &          $\sim$2.5   & $512^3$    & 1.00     & 1.15 \\
G1-B2        &   7.9  &  0.396 &   2    & 0.792   &         $\sim$2.5   & $512^3$    & 0.92     & 0.86  \\
G1-B4        &  7.7   &  0.191  &   4  & 0.766  &          $\sim$2.5   & $512^3$    & 0.90      & 0.72 \\ 
\hline
G2-B0.03        &   12   &  40.5  &  0.03  &    1.23 &           $\sim$10   & $512^3$     & 2.00     & 1.95  \\
G2-B0.1        &   12   &  11.7   &   0.1  &  1.17  &               $\sim$10   & $512^3$    & 1.79     & 1.77  \\
G2-B0.2        &  11   &  5.62   &   0.2  &   1.12  &               $\sim$10   & $512^3$    & 1.61     & 1.62  \\
G2-B0.4        &   11   &  2.67    &   0.4  &   1.07  &               $\sim$10   & $512^3$    & 1.30     & 1.41 \\
G2-B0.6        &  11  &   1.75   &   0.6  & 1.05  &               $\sim$10   & $512^3$    & 1.14     & 1.30  \\
G2-B1        &   10   &   1.04   &  1  &  1.04    &              $\sim$10   & $512^3$    &   0.98     & 1.21  \\
G2-B2        &  10   &   0.506  &  2  & 1.01     &               $\sim$10   & $512^3$    &    0.89     & 1.05  \\
G2-B4        &   9.9   &  0.248  & 4   &    0.991  &                $\sim$10  & $512^3$    & 0.86      & 0.86 \\ 
\hline
G3-B0.01        &  3.9   &  125   &  0.01   & 1.25  &      $\sim$10   & $512^3$    & 2.17     & 2.15  \\
G3-B0.03        &   3.9   &  41.0    &  0.03   &  1.23 &    $\sim$10   & $512^3$    & 2.13     & 2.14  \\
G3-B0.1      &   3.7   &  11.7    &  0.1  & 1.17  &         $\sim$10   & $512^3$    & 1.96     & 1.92 \\
G3-B0.2        &   3.5   & 5.58   & 0.2   &  1.12 &         $\sim$10   & $512^3$    & 1.72     & 1.78  \\
G3-B0.5        &   3.3   &  2.10   &  0.5   & 1.05  &        $\sim$10   & $512^3$    & 1.36     & 1.54  \\
G3-B1        &   3.3   &  1.06   &    1   & 1.06 &          $\sim$10   & $512^3$    & 1.08     & 1.34  \\
G3-B2        &  3.3   &  0.520  &   2   & 1.04 &           $\sim$10  & $512^3$    & 0.99      & 1.18 \\ 
\enddata
\tablecomments{
\tablenotetext{a}{The number after `B' actually denotes the Alfv\'en speed of the mean magnetic field.} 
\tablenotetext{b}{The average sonic Mach number after saturation.} 
\tablenotetext{c}{The average Alfv\'enic Mach number ($M_A =  v_{rms}/V_{A0}$) after saturation.} 
\tablenotetext{d}{The Alfv\'en speed of the mean magnetic field. $V_{A0}=B_{0}/\sqrt{4 \pi \bar{\rho} }$. }
\tablenotetext{e}{The size of the computational box divided by the driving scale. Note that $L/L_{f}=k_f$.} 
\tablenotetext{f}{The average small-scale E/B power asymmetry after saturation. The line of sight (LOS) is perpendicular to ${\bf B}_0$.} 
\tablenotetext{g}{The average small-scale E/B power asymmetry after saturation.  The angle between the LOS and ${\bf B}_0$ is $\sim$30$^\circ$. 
To be exact, the angle is $\tan^{-1}0.5=26.6^\circ$.} 
 }
\label{tbl:sim}
\end{deluxetable*}

\begin{figure*}
\centering
\includegraphics[width=0.3\textwidth, trim=50 158 250 290, clip]{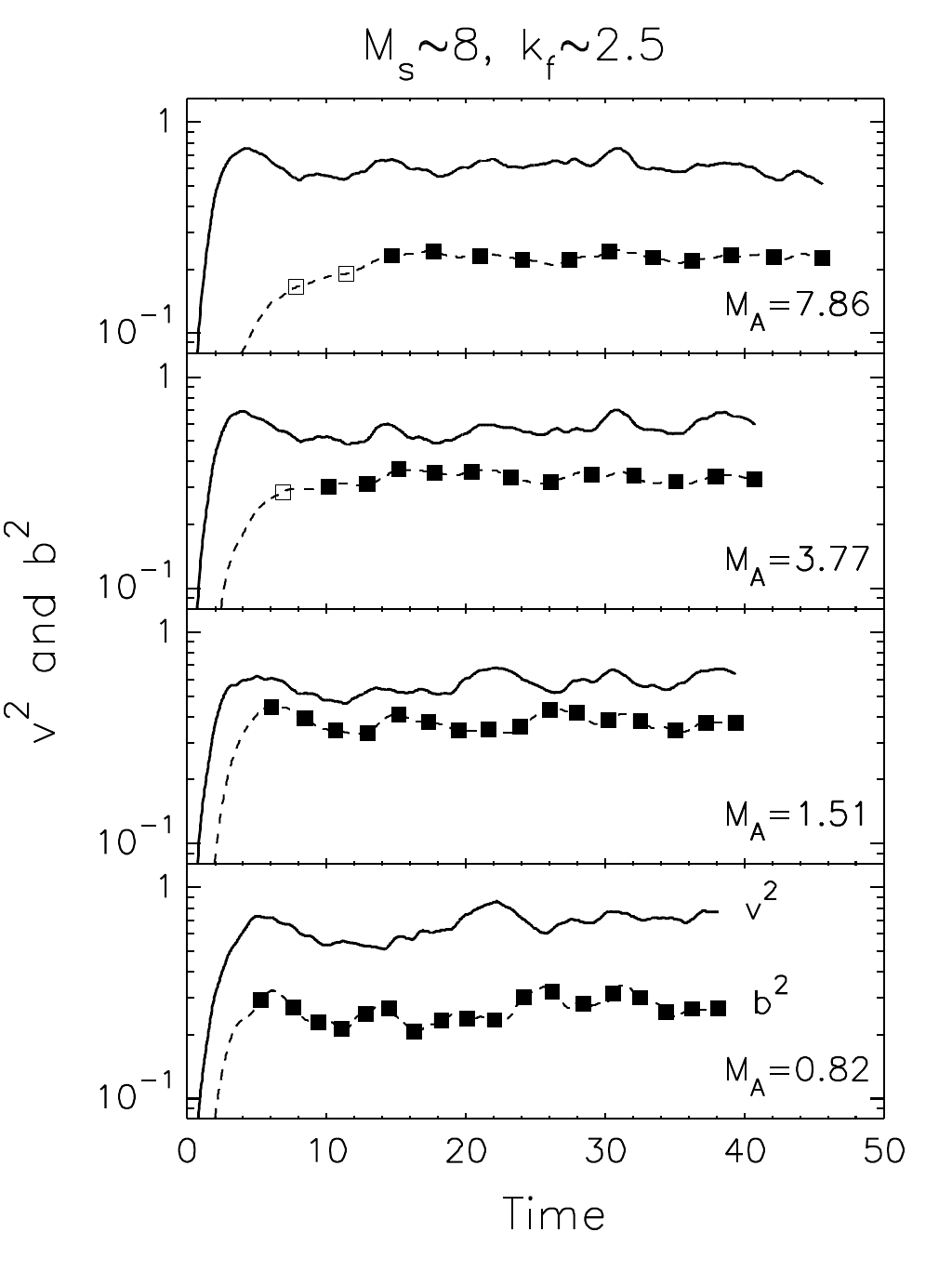}
\includegraphics[width=0.3\textwidth, trim=50 158 250 290, clip]{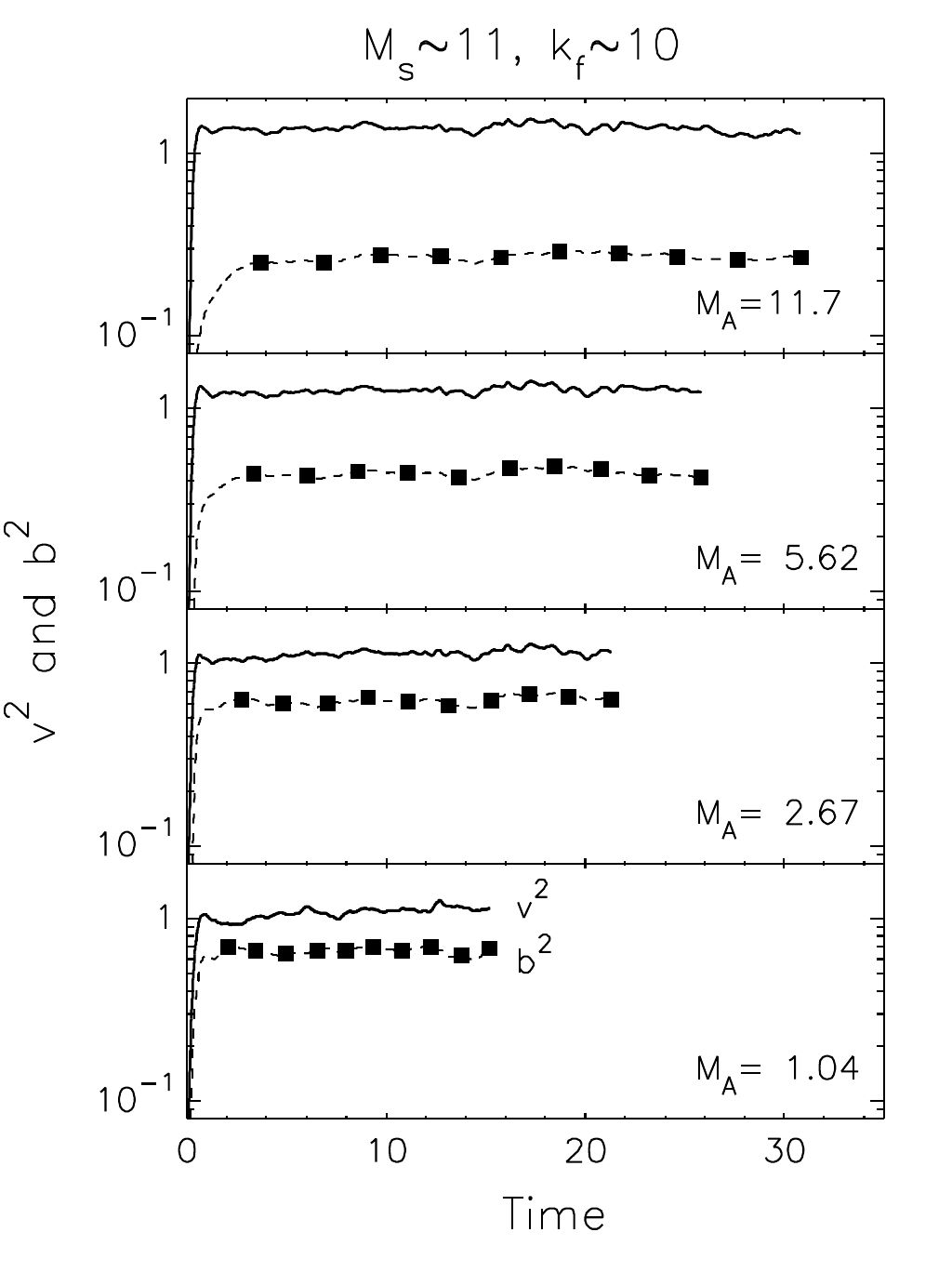}
\includegraphics[width=0.3\textwidth, trim=50 158 250 290, clip]{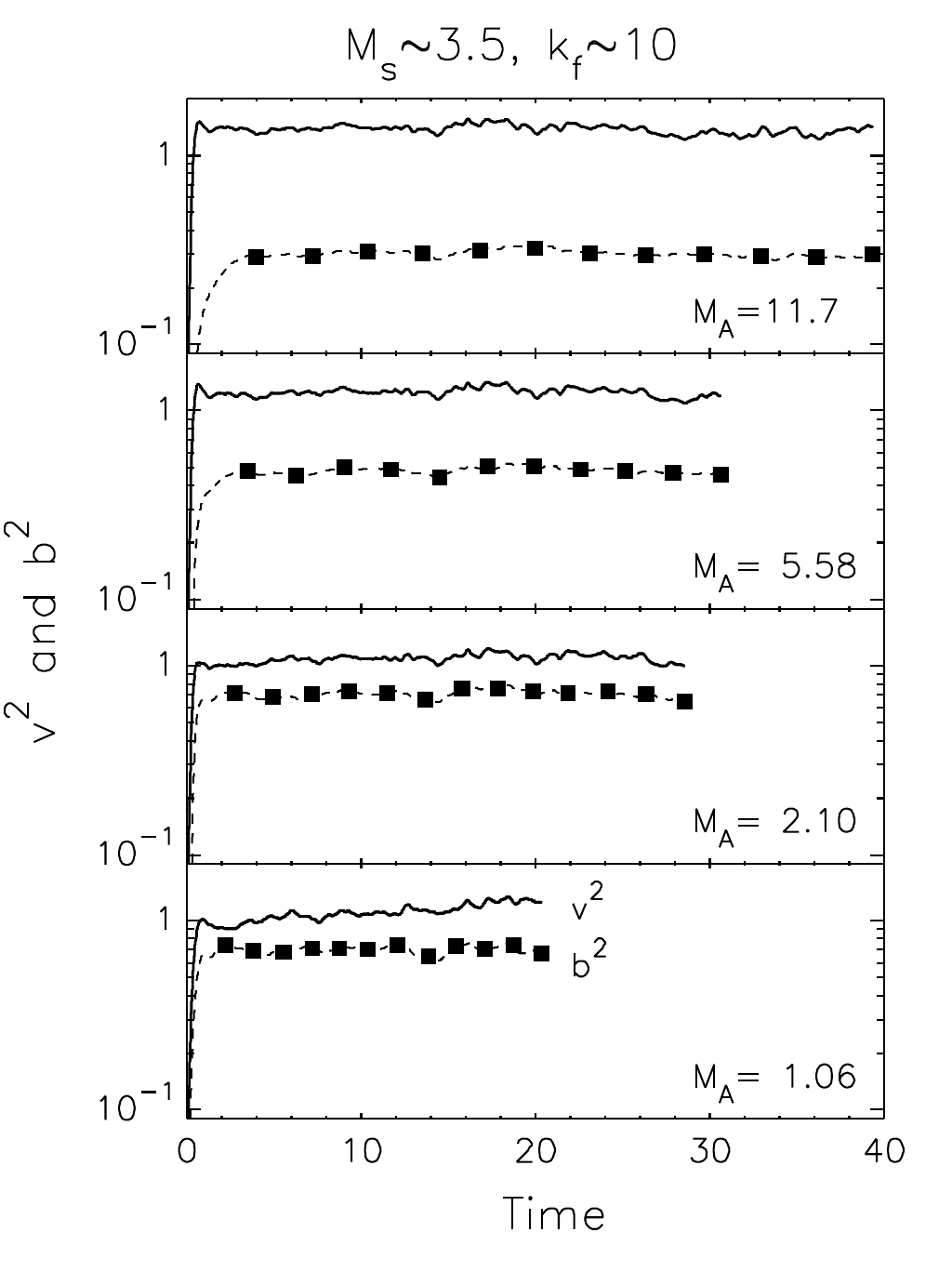}
\caption{Time evolution of $v^2$ and $b^2$ for selected runs in simulation groups {\bf G1} (left panels), {\bf G2} (middle panels), and
{\bf G3} (right panels).
From top to bottom, the Alfv\'enic Mach number $M_A$   ($=\sqrt{ 4 \pi \bar{\rho} }~  v_{rms} /{B_0}$)  
decreases.
Left: From top to bottom, panels correspond to Runs G1-B0.1, G1-B0.2,  G1-B0.5, and G1-B1, respectively.
Middle: From top to bottom, panels correspond to Runs G2-B0.1, G2-B0.2,  G2-B0.4, and G2-B1, respectively.
Right: From top to bottom, panels correspond to Runs G3-B0.1, G3-B0.2,  G3-B0.5, and G3-B1, respectively.}
\label{fig:en}
\end{figure*}


\section{Results} \label{sect:result}
\subsection{Time evolution of $v^2$ and $b^2$}  
Figure \ref{fig:en} shows time evolution of $v^2$ and $b^2$.
The left, middle, and right panels show selected  runs from simulation groups {\bf G1}, {\bf G2} and {\bf G3}, respectively.
From top to bottom, the strength of the mean magnetic field ($B_0$) becomes stronger.
Sold and dashed lines denote $v^2$ and $b^2$, respectively.
As turbulence develops, fluctuating magnetic field initially grows and then reaches statistically-stationary saturation state. 
The growth time is longer when the mean field is weaker.
In this paper, we analyze the E/B power ratio after turbulence reaches saturation state.
The filled symbols denote the data we use for our analysis.

\begin{figure*}
\centering
\includegraphics[scale=0.4]{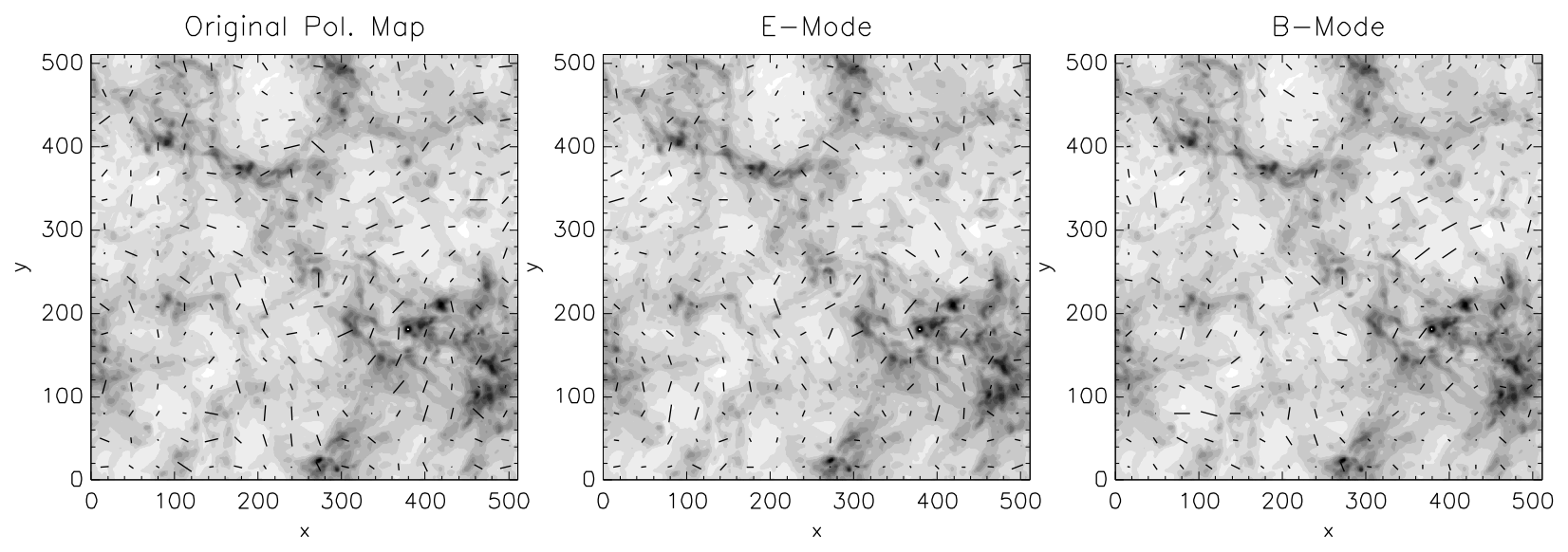}
\caption{ Polarization maps for Run G1-B0.1, in which the Alfv\'en Mach number is $\sim 8$ (see Table 1 and the top-left panel of Figure \ref{fig:en}).
 We separate the original polarization pattern into E (gradient-type) and B (curl-type) modes. 
Left: Original polarization map.
   Middle: E modes.  
   Right: B modes. The direction of the mean magnetic field is along the $x$-axis and perpendicular to the line of sight (LOS).}
\label{fig:conto}
\end{figure*}

\subsection{E/B power asymmetry in the case the LOS is perpendicular to ${\bf B}_0$} \label{sect:90}
In this subsection, we assume the direction of the mean magnetic field is perpendicular to the LOS and
study the relation between the E/B power asymmetry and the Alfv\'en Mach number $M_A$.
The coordinate system we
adopted in this subsection is such that the LOS is along  the $z$-axis and the sky plane coincides with the $xy$-plane. 
The mean magnetic field lies in the sky plane and is along the horizontal axis (i.e., $x$-axis).

Following the procedure in \S\ref{sect:method}, we obtain synthetic polarization maps and
separate E and B modes in Fourier space.
 Figure \ref{fig:conto} shows polarization maps for Run G1-B0.1,  which  is a supersonic and super-Alfv\'enic turbulence.
The left panel is the original polarization map, and the middle and the right panels are maps for E and B modes, respectively.
As explained in the previous paragraph, the LOS is along the $z$-axis and the direction of the mean magnetic field is along the $x$-axis.
Since the mean magnetic field is weak, the direction of polarization shows large variations.

\begin{figure*}
\centering
\includegraphics[width=0.3\textwidth, trim=50 158 250 290, clip]{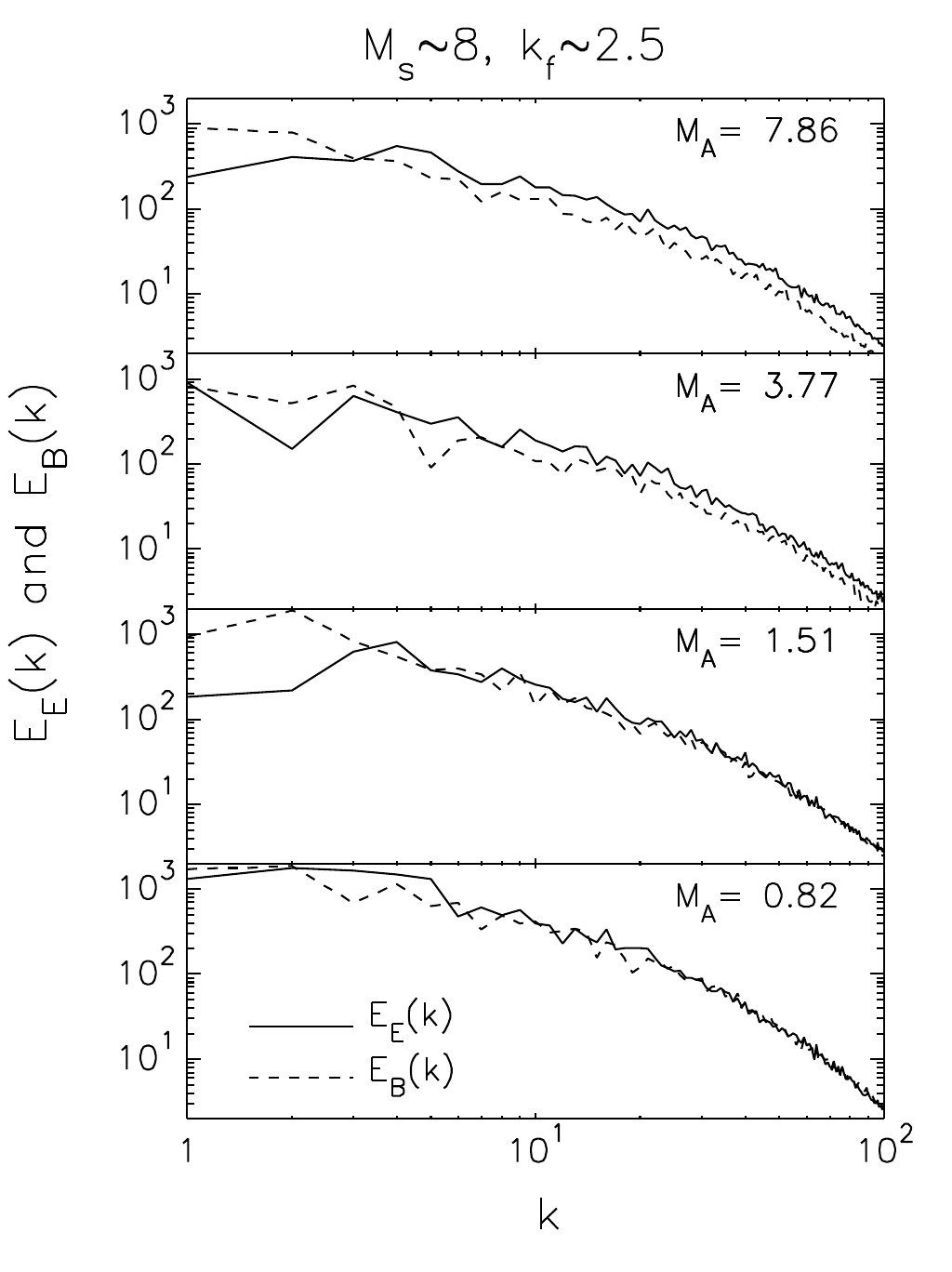 }
\includegraphics[width=0.3\textwidth, trim=50 158 250 290, clip]{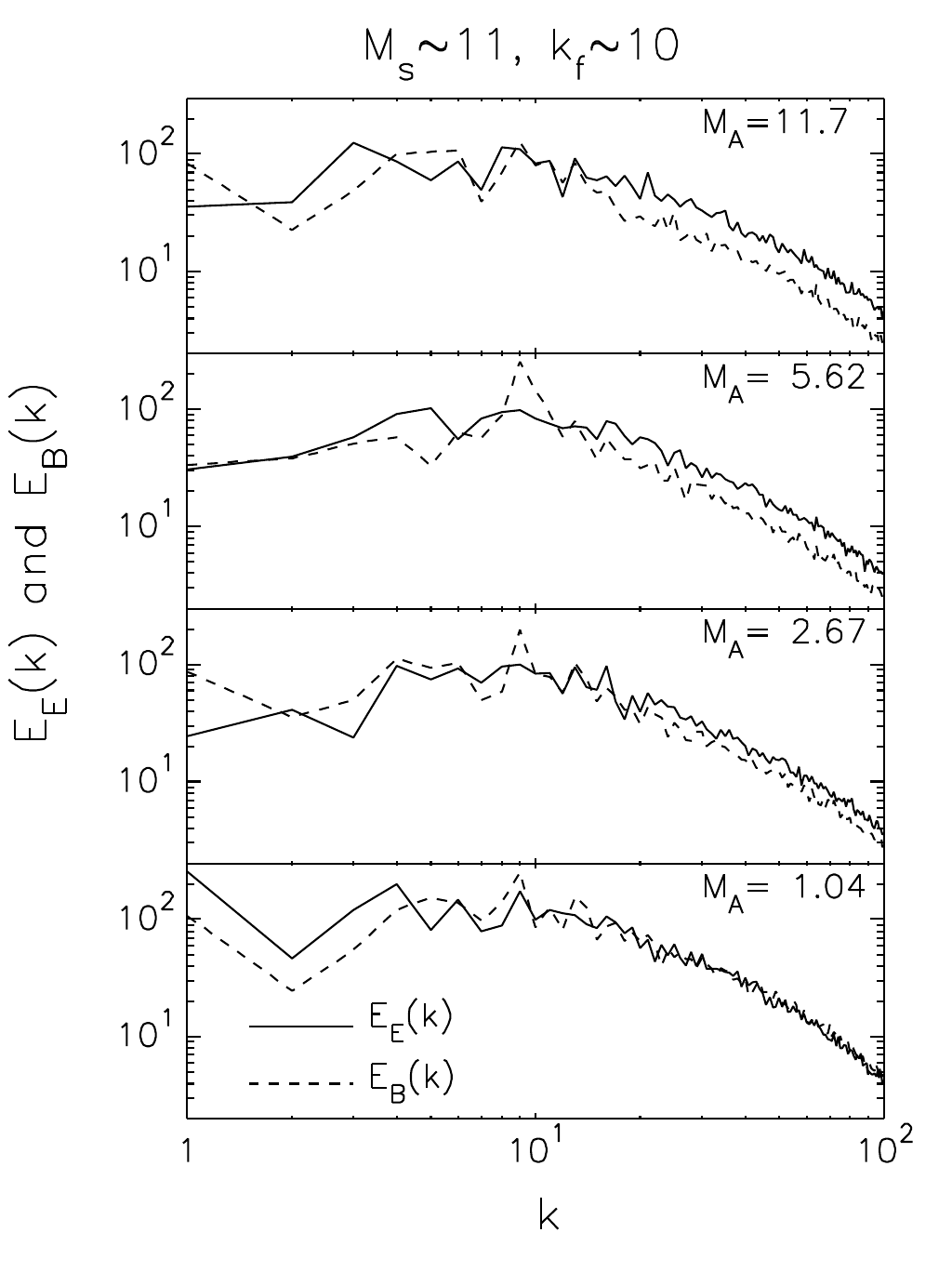}
\includegraphics[width=0.3\textwidth, trim=50 158 250 290, clip]{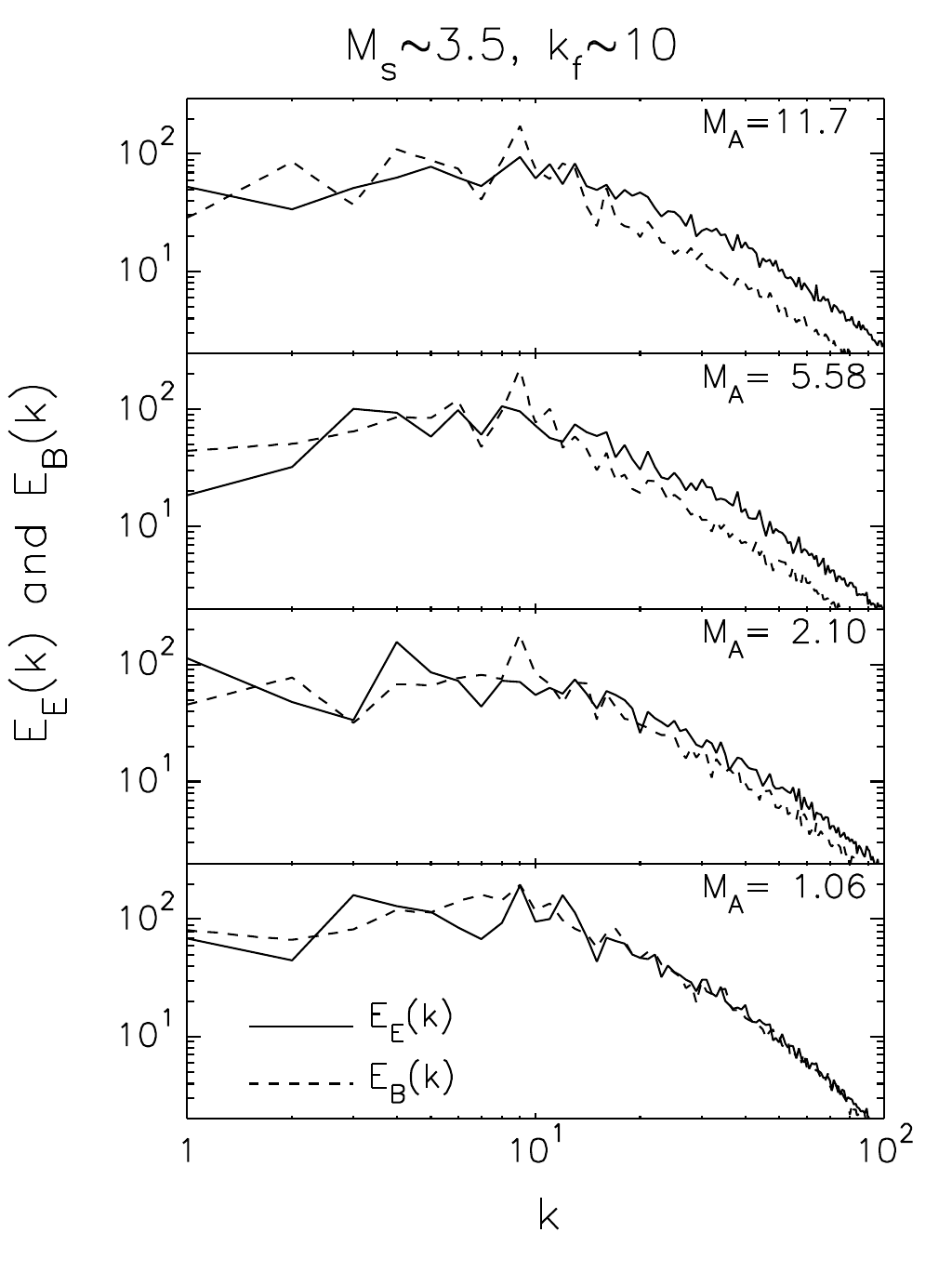}
\caption{
Energy spectra of E and B modes.
 In all panels, sold and dashed curves denote E and B-mode spectra, respectively. 
When Alfv\'enic Mach number $M_A$ is large, the E/B power asymmetry is large  for $k \gtrsim 20$ (see top panels).
The E/B power asymmetry gradually decreases as $M_A$ decreases.
Runs shown here are the same as Figure \ref{fig:en}.
Left: Spectra for simulation group {\bf G1}.  
  Middle: Spectra for simulation group {\bf G2}.
   Right: Spectra for simulation group  {\bf G3}.
The direction of the mean magnetic field is perpendicular to the LOS. }
\label{fig:sp}
\end{figure*}

\begin{figure*}
\centering
\includegraphics[width=0.3\textwidth, trim=50 158 210 290, clip]{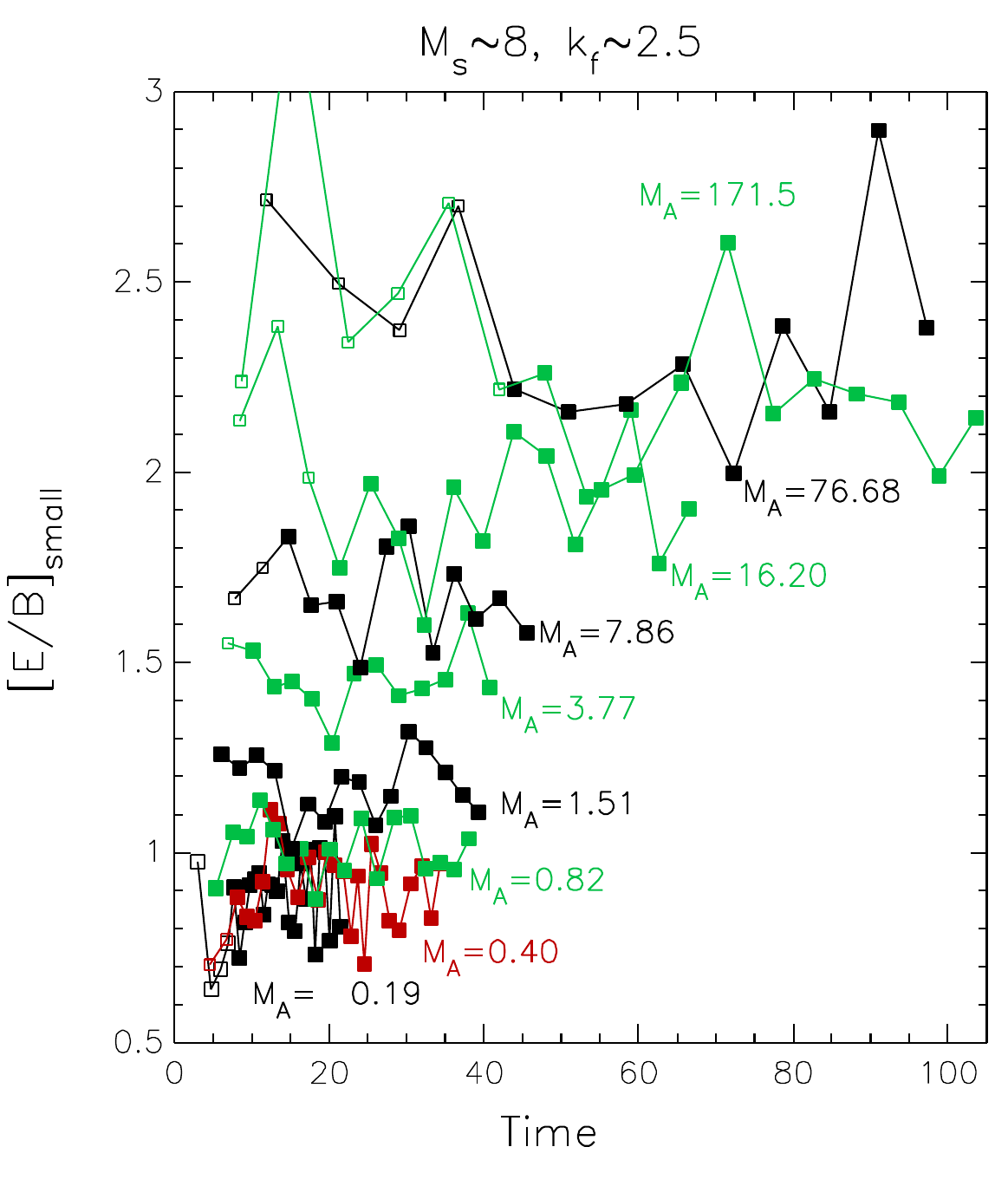}
\includegraphics[width=0.3\textwidth, trim=50 158 210 290, clip]{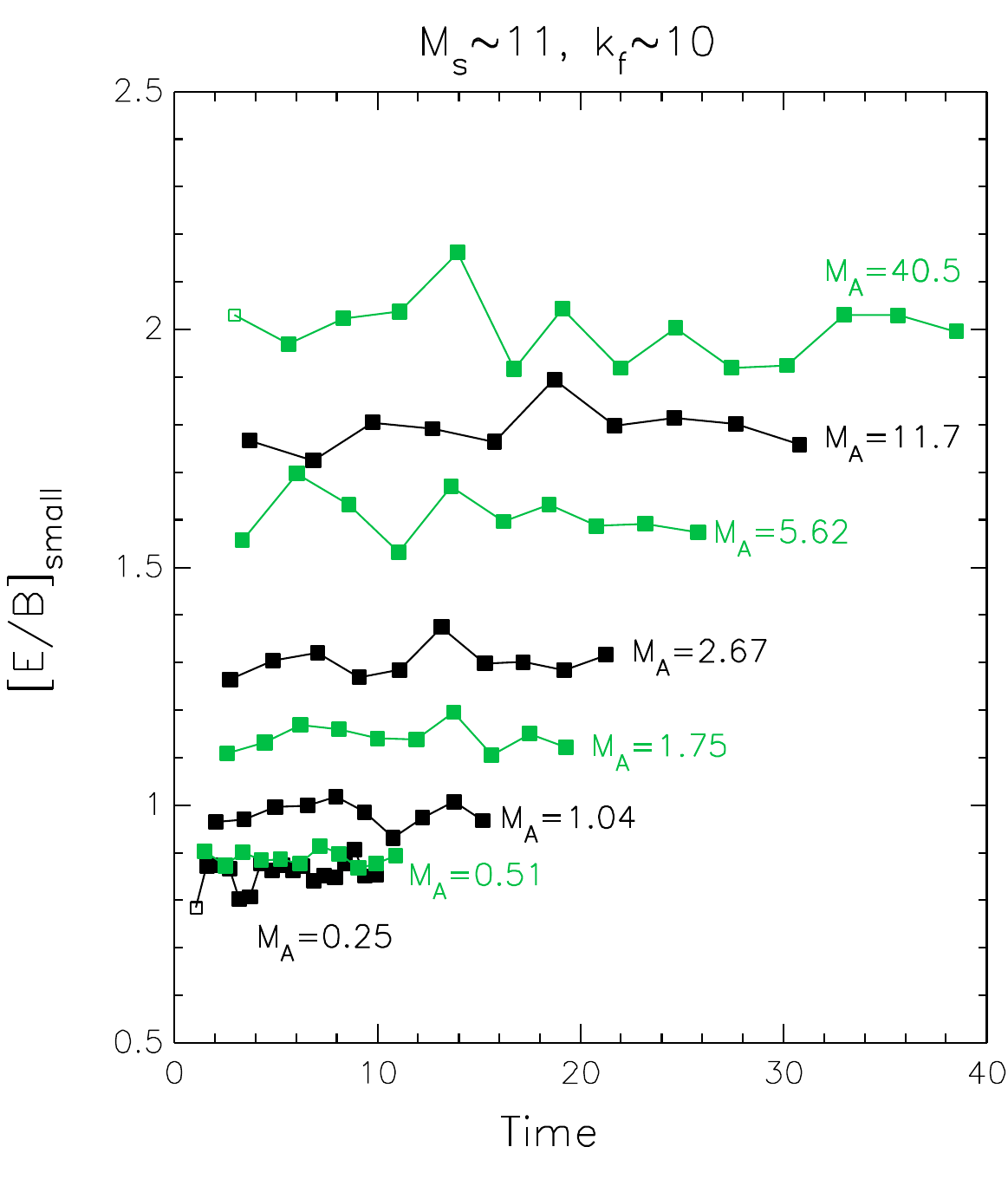}
\includegraphics[width=0.3\textwidth, trim=50 158 210 290, clip]{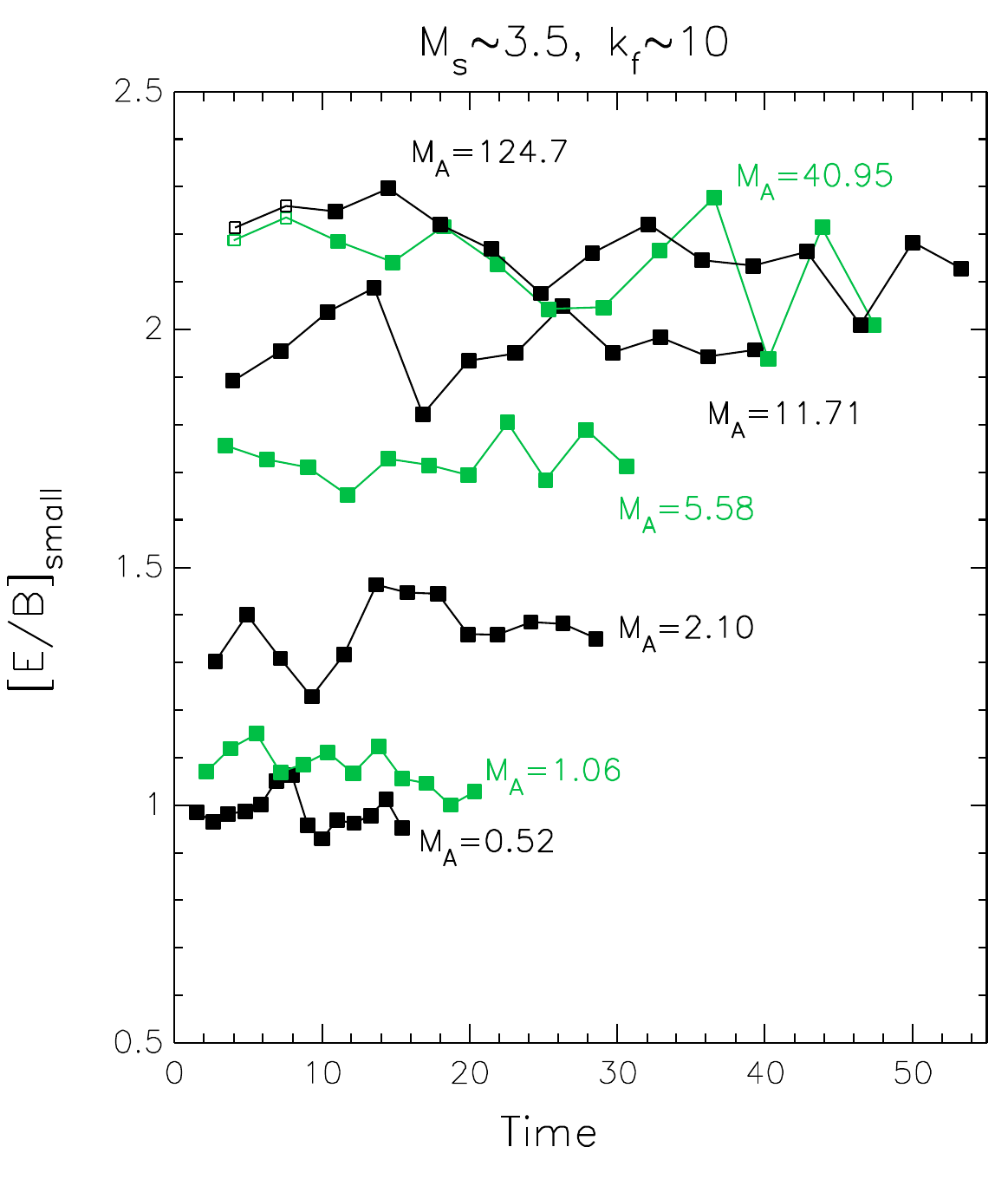}
\caption{Small-scale E/B power asymmetry vs. time.
 All runs listed in Table 1 are shown.
 Left: Simulation group {\bf G1}. 
 Middle: Simulation group {\bf G2}. 
 Right: Simulation group {\bf G3}.
The direction of the mean magnetic field is perpendicular to the LOS. }
\label{fig:ratio}
\end{figure*}


From the E and B modes obtained in Fourier space, we calculate spectra of them.
Figure \ref{fig:sp} shows E and B-mode spectra for the runs shown in Figure \ref{fig:en}.
As in Figure \ref{fig:en}, the left, middle, and right panels correspond to simulation groups {\bf G1}, {\bf G2} and {\bf G3}, respectively, and the mean field becomes stronger as we move from top to bottom.
The solid and dashed curves denote E and B-mode spectra, respectively.
As we can see in the figure, E-mode spectra are clearly larger than B-mode spectra (i.e.,  $E_E(k)>E_B(k)$)
for $k \gtrsim 20$ in top panels, while $E_E(k) \sim E_B(k)$ in bottom panels.
That is, the E/B power ratios (i.e., the values of $E_E(k)/E_B(k)$; see Equation (\ref{eq:eb_asym}) for definition) are clearly larger than 1 on top panels and they approach $\sim1$ as we move down to bottom panels.
Note that the Alfv\'en Mach number $M_A$ is $\sim 10$ in top panels and it gradually decreases as we move down to bottom panels.
In bottom panels, $M_A \sim 1$.

In Figure \ref{fig:ratio}, we plot time evolution of small-scale E/B power ratio $[E/B]_{small}$.
We show results for all simulations listed in Table 1. 
The left, middle, and right panels correspond to simulation groups {\bf G1}, {\bf G2} and {\bf G3}, respectively.
Filled symbols denote data after turbulence reaches saturation.
We write the Alfv\'en Mach number $M_A$ of a simulation next to the line that represents the simulation.
All 3 panels in the figure show a similar trend: When $M_A$ is large, the E/B power ratio is large and, as $M_A$ decreases, the ratio decreases.

\begin{figure*}
\includegraphics[width=0.45\textwidth, trim=50 180 200 360, clip]{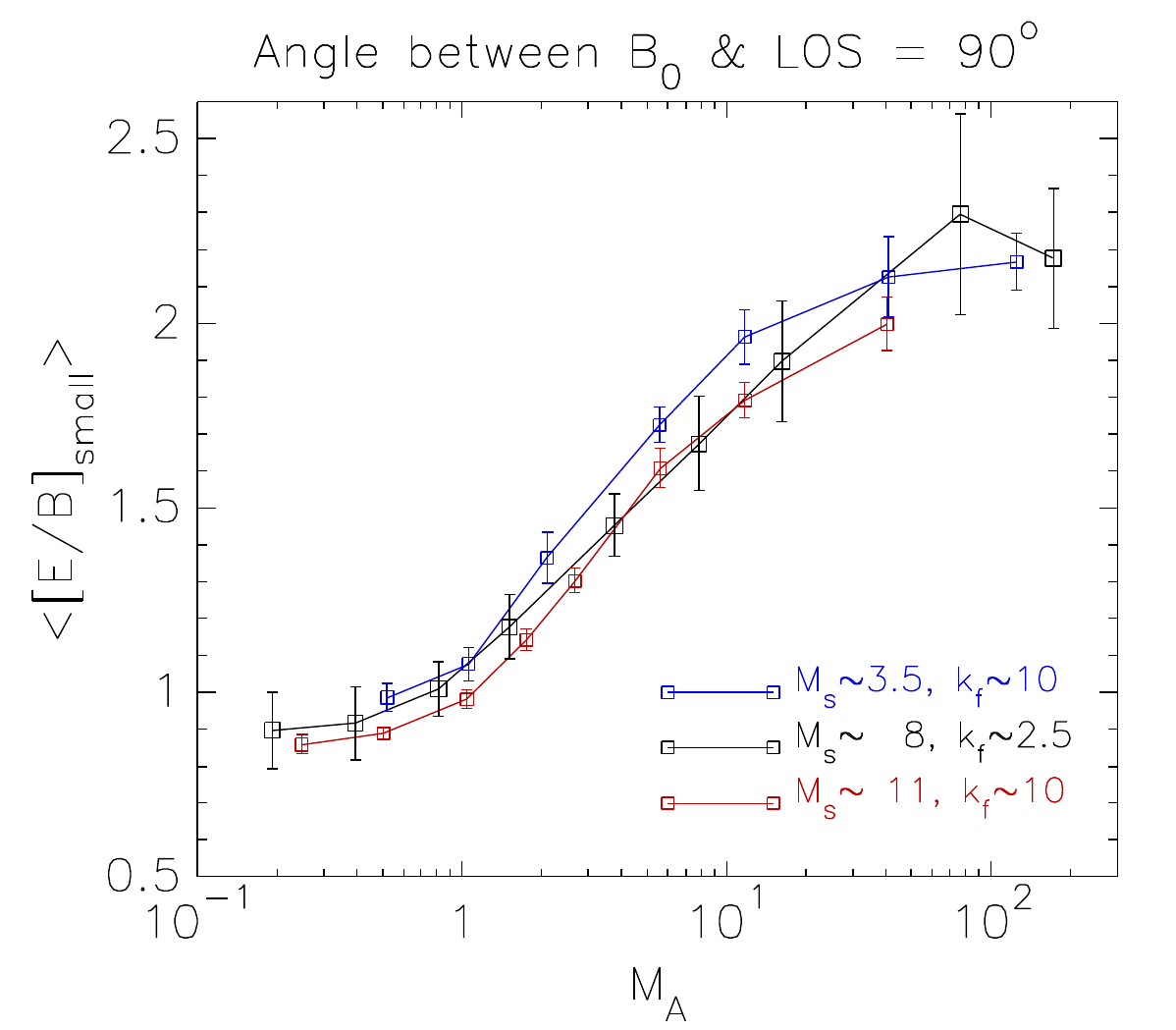}
\includegraphics[width=0.45\textwidth, trim=50 180 200 360, clip]{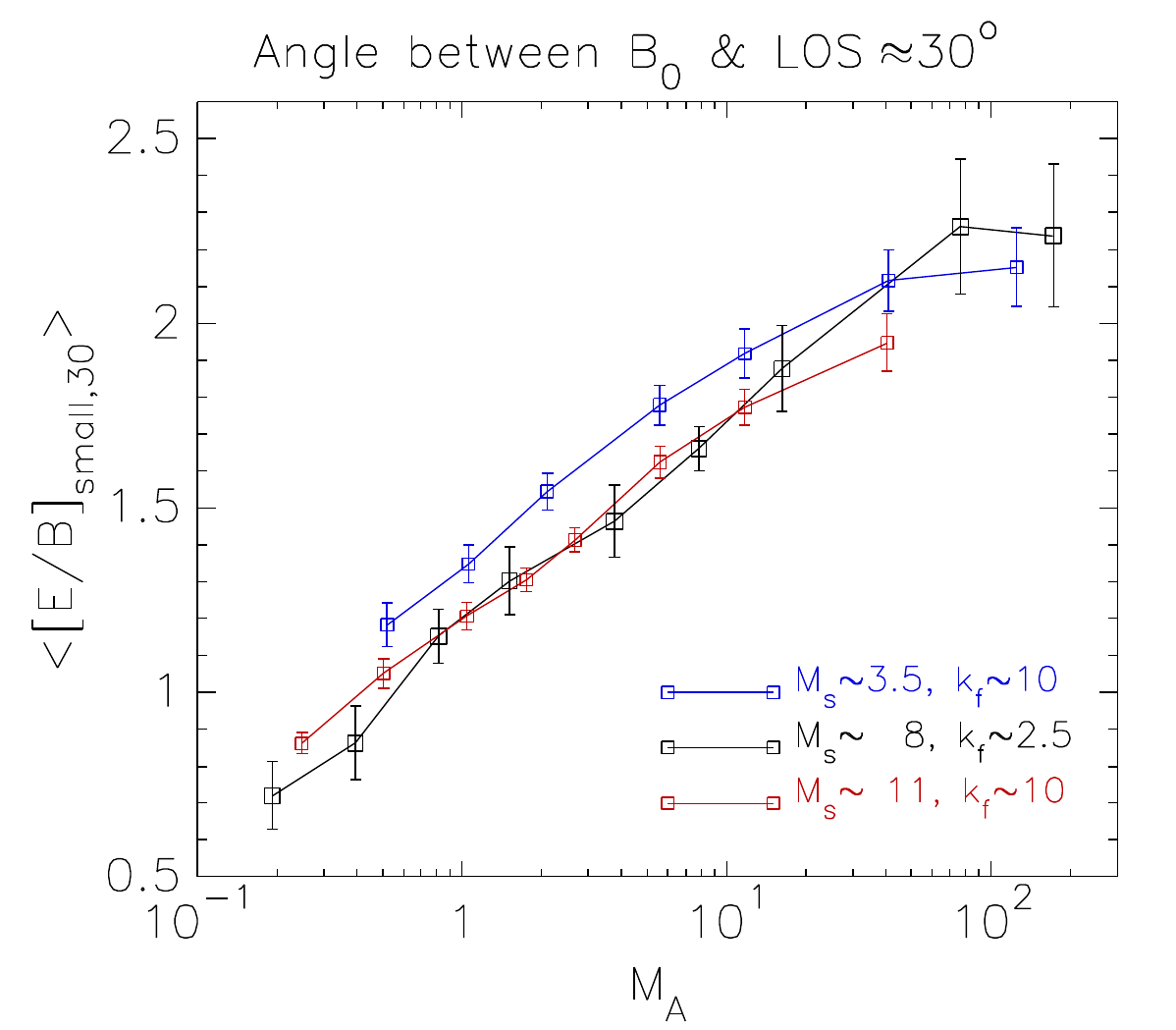}
\caption{Small-scale E/B power asymmetry vs. $M_A$.
Left: The direction of the mean magnetic field is perpendicular to the LOS.
Right: The LOS make an angle of $\sim$30$^\circ$ with the mean magnetic field.
To be exact, the angle is $\tan^{-1}0.5=26.6^\circ$.}  
\label{fig:sum}
\end{figure*}

The left panel of Figure \ref{fig:sum} shows average values of $[E/B]_{small}$ vs. $M_A$
in the case the LOS is perpendicular to the mean magnetic field.
The average is taken after turbulence reaches saturation (see the filled symbols in Figure \ref{fig:ratio}).
We distinguish different simulation groups with different colors: the black, red and blue lines denote
simulation groups {\bf G1}, {\bf G2}, and {\bf G3}, respectively.
We notice that the average E/B power ratio $<[E/B]_{small}>$ is a steep increasing function of $M_A$ for $1 \lesssim M_A \lesssim 30$, which means that
the E/B power asymmetry could be used to determine $M_A$ when $M_A$ lies in the range $1 \lesssim M_A \lesssim 30$.
When $M_A \lesssim 1$, it seems that $<[E/B]_{small}>$ shows a very weak dependence on $M_A$.
Therefore, it may be difficult to determine $M_A$ using the E/B power asymmetry when $M_A \lesssim 1$.
When  $M_A \gtrsim 30$, it seems that  $<[E/B]_{small}>$  levels off, which implies that
the E/B power asymmetry may not be used to constrain the value of $M_A$ for $M_A \gtrsim 30$.
Note however that it is highly unlikely that turbulence in molecular clouds has an $M_A$ larger than $\sim 30$.
Roughly speaking, we can use the following formula to fit the data in the left panel of Figure \ref{fig:sum}:
\begin{equation}   \label{eq:90}
 <[E/B]_{small}> = \begin{cases}
              \lesssim 1                  &  \text{if } M_A \lesssim 1, \\
              0.75 \log_{10} M_A +1     &  \text{if } 1 \lesssim M_A \lesssim 30, \\
              \sim 2.1                     &   \text{if }   M_A \gtrsim 30.
              \end{cases}
\end{equation}

The left panel of Figure \ref{fig:sum} shows that $<[E/B]_{small}>$ has a weak dependence on the sonic Mach number.
If we compare  the blue ($M_s \sim 3.5$) and the red ($M_s \sim 11$) lines, $<[E/B]_{small}>$  for the blue line is systematically higher than that for the red line.
Note however that the difference is small.
The black line, which is  for an intermediate sonic Mach number of $M_s \sim 8$, lies between the blue and the red lines.

\subsection{E/B power asymmetry in the case the LOS makes an angle of $\sim$30$^\circ$ with ${\bf B}_0$}  \label{sect:30}
In \S\ref{sect:90}, we have assumed that the LOS is perpendicular to the mean magnetic field.
To see the effect of the viewing angle, we repeat the calculation for the case that the LOS makes an angle of $\sim$30$^\circ$ with ${\bf B}_0$.

The right panel of Figure \ref{fig:sum} shows average values of $[E/B]_{small}$ vs. $M_A$
in the case the LOS makes an angle of $\sim$30$^\circ$ with ${\bf B}_0$.
If we compare  the left and the right panels of Figure \ref{fig:sum}, we do not see a significant difference.
Note however that the values of the average E/B power ratio in the right panel are slightly higher than those in the left panel
especially for $M_A \lesssim 3$, while
the values in both panels are similar  for $M_A \gtrsim 3$.
It is interesting that the lines in the right panel continue to decrease when $M_A$ drops below unity, which is not seen in the left panel.

We may use the following formula to fit the data in the right panel of Figure \ref{fig:sum}:
\begin{equation}   \label{eq:30}
 <[E/B]_{small,30}> = \begin{cases}
              0.6 \log_{10} M_A +1.2     &  \text{if } M_A \lesssim 30, \\
              \sim 2.1                     &   \text{if }   M_A \gtrsim 30.
              \end{cases}
\end{equation}
Note however that the behavior of $<[E/B]_{small,30}>$ below $M_A \sim 1$ is very uncertain.

\subsection{Total E/B power asymmetry}

\begin{figure*}
\includegraphics[width=0.45\textwidth, trim=50 180 200 360, clip]{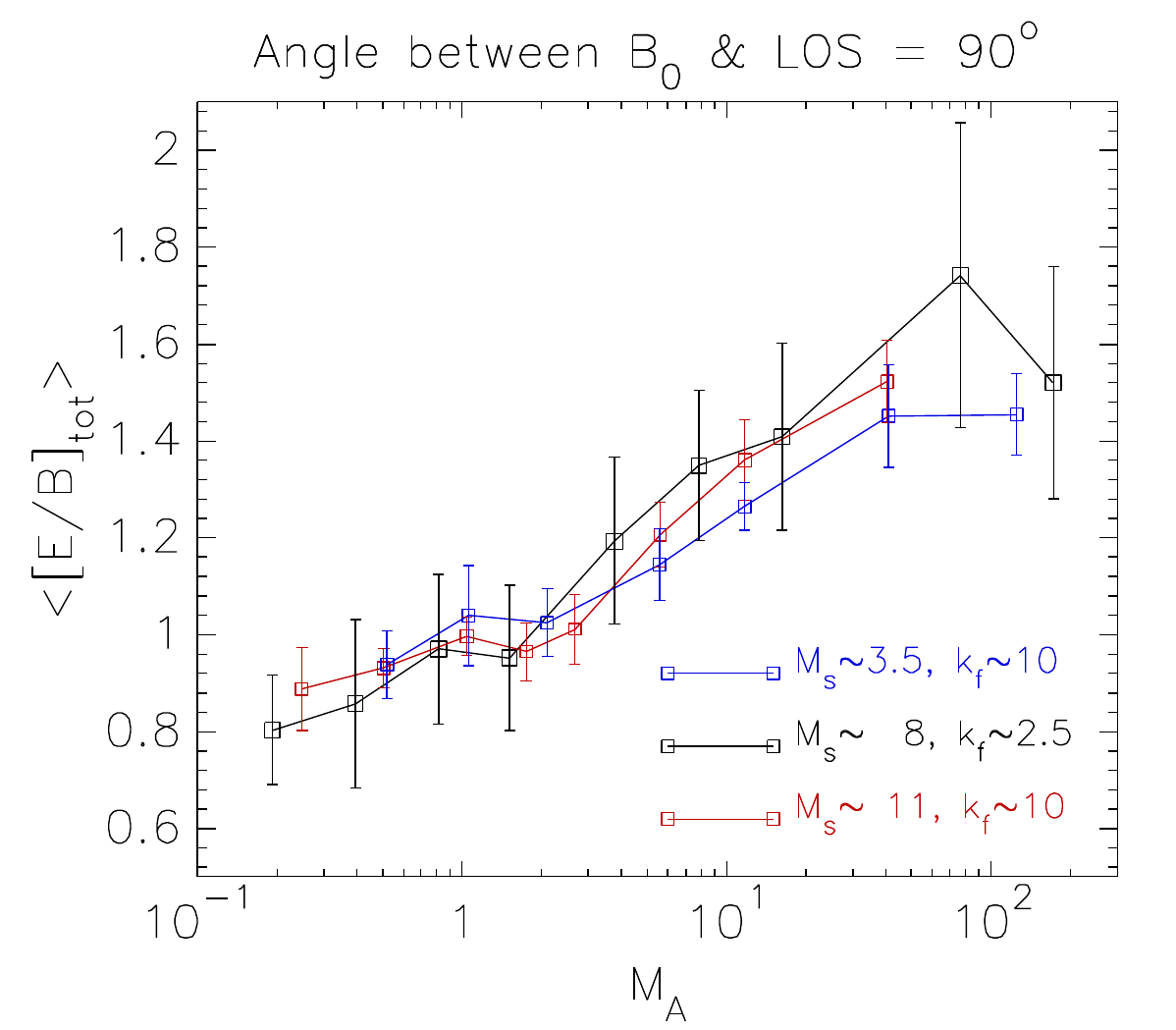}
\includegraphics[width=0.45\textwidth, trim=50 180 200 360, clip]{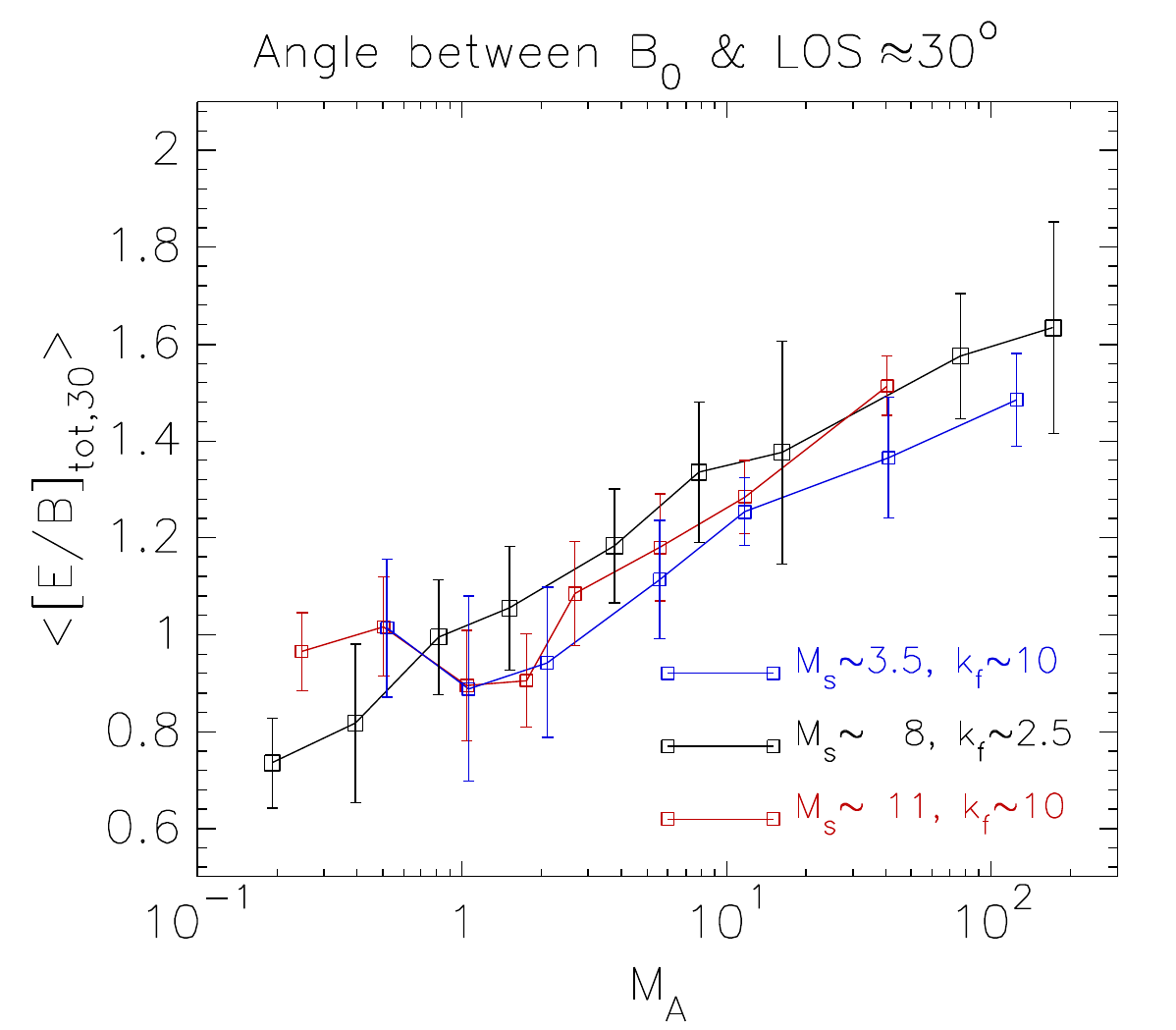}
\caption{Total E/B power asymmetry vs. $M_A$.
Left: The direction of the mean magnetic field is perpendicular to the LOS.
Right: The LOS make an angle of $\sim$30$^\circ$ with the mean magnetic field.
To be exact, the angle is $\tan^{-1}0.5=26.6^\circ$.}   
\label{fig:tot}
\end{figure*}

We have considered the E/B power asymmetry on small spatial scales.
We can also  calculate the asymmetry using the total powers of E and B modes:
\begin{equation}
  [E/B]_{tot} = \frac{ \int_1^{k_{max}} E_E(k) dk }{  \int^{k_{max}}_{1} E_B(k) dk} .   \label{eq:eb_tot}
\end{equation}
Figure \ref{fig:tot} shows the results.
The observing geometry is the same as Figure \ref{fig:sum}.
Total powers of E and B modes also show that the E/B power asymmetry increases as $M_A$ increases,
However, the relation between the asymmetry and $M_A$ in Figure \ref{fig:tot} is not as tight as that in Figure \ref{fig:sum}.

\section{Discussion and Summary}   \label{sect:disc}

In this paper, we have investigated the relation between the E/B power asymmetry arising from magnetically aligned dust grains and $M_A$ in supersonic turbulence, which is relevant to molecular clouds.
We have found that the E/B power asymmetry increases as $M_A$ increases.
At this moment, it is not clear why the E/B power asymmetry shows such a behavior.

\begin{figure}
\includegraphics[width=0.45\textwidth, trim=50 180 200 360, clip]{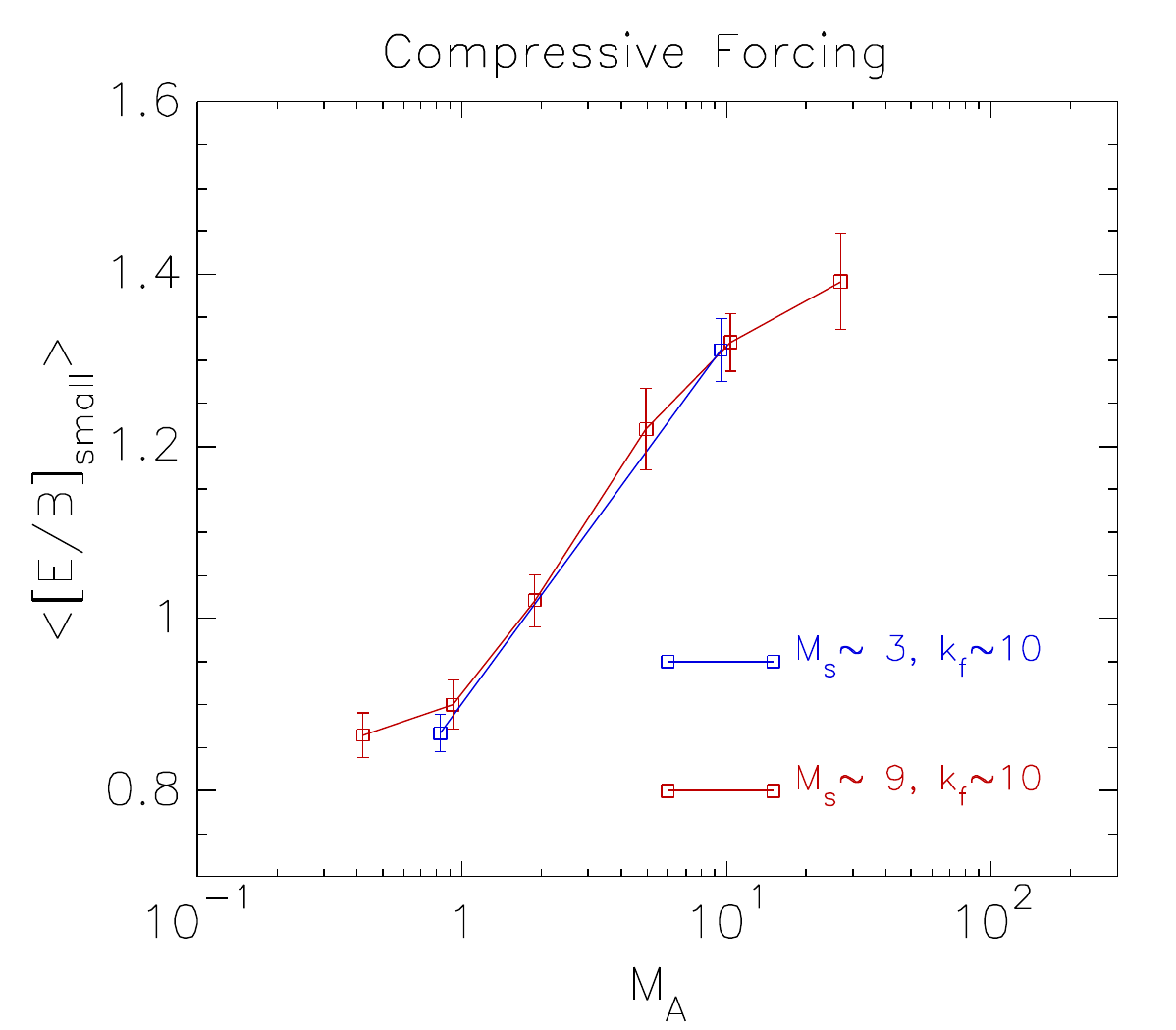}
\caption{Small-scale E/B power asymmetry for compressively-driven turbulence. The direction of the mean magnetic field is perpendicular to the LOS.}
\label{fig:comp}
\end{figure}

Our study has several limitations. 
Presumably the most important limitation would be the driving mode of turbulence.
In this study, we have considered purely solenoidal driving only.
However, realistic astrophysical driving, such as stellar feedback on small scales and supernova feedback on large scales, can have substantial amount of compressive components (see, for example, Kainulainen and Federrath, 2017).
Earlier studies showed that turbulence driving mode has significant impacts on density fluctuations (Federrath et al. 2010)  and magnetic field generations (see, for example, Lim et al. 2020), which may affect the E/B ratio.
Therefore, we perform numerical simulations with compressive driving and check if both solenoidal and compressive forces produce similar E/B ratios. We show our preliminary results in Figure \ref{fig:comp}. 
The red and blue lines denote the E/B ratio for $M_s \sim 9$ and $\sim 3$, respectively.
The simulation setups for compressively-driven turbulence are virtually identical to those for simulation groups {\bf G2} and {\bf G3}, except for the driving mode. 
We first notice that the small-scale E/B power asymmetry for compressively-driven turbulence also increases 
as $M_A$ increases for  $1 \lesssim M_A \lesssim 30$.
We still observe that the E/B power asymmetry is greater than or similar to 1 when $M_A  \gtrsim 1$.
Note however that the E/B power asymmetry for compressively-driven turbulence (Figure \ref{fig:comp}) shows smaller variation than the one for solenoidally-driven turbulence (see the left panel of Figure \ref{fig:sum}). 
Therefore, it will be important to know the driving mode of turbulence in order to accurately constrain $M_A$ from the E/B ratio.
Note that the DCF method is also affected by turbulence driving mode (Yoon \& Cho 2019).

Our study has other limitations.
First, we have considered supersonic turbulence only.
The range of the sonic Mach number we have considered is $3 \lesssim M_s \lesssim 11$.
Therefore, our study is relevant mainly to diffuse molecular clouds in the ISM.
Second, we have not included the effect of self-gravity.
Since, self-gravity induces compressive motions, it may also affect the E/B power asymmetry ratio.
Since self-gravity is not included in our simulations, our results are relevant to diffuse molecular clouds, rather than proto-stellar cores.
Third, we have not considered realistic astrophysical driving, such as stellar winds, jets, and supernova explosions.
In reality, it is also possible that small-scale driving, such as stellar feedback, takes place at high-density and large-scale energy is injected at low-density.
However, we have not included these factors: we simply drive turbulence in Fourier space by stochastically changing velocity amplitudes.
We will present more comprehensive study covering more parameter space  and including more realistic driving
elsewhere.

Our results suggest that the E/B power asymmetry can be used to constrain $M_A$ in super-Alfv\'enic turbulence.
More precisely speaking, it can be used to determine $M_A$ for $1\lesssim M_A \lesssim 30$.
The ratio can also tell us whether $M_A$ is larger than $\sim 30$ or less than $\sim 1$.
Since 
\begin{equation}
   M_A = \sqrt{ 4 \pi \bar{\rho}  } v_{rms}/B_0,
\end{equation}
we can easily obtain $B_0$ from $M_A$ if we know $v_{rms}$ and $\bar{\rho}$.

The use of the E/B power asymmetry has following advantages.
First, it can cover a parameter space which may not be reached by the DCF method.
While the DCF method works mainly for $M_A \lesssim 1$,  the E/B power asymmetry method works for $M_A \gtrsim 1$.
In this sense, the E/B power asymmetry method is complementary to the DCF method.
Second, the E/B power asymmetry is not affected by the number of independent eddies along the LOS.
Note that  the DCF method does not work correctly if there are many independent eddies along the LOS (see discussions in 
Myers \& Goodman 1991;
Zweibel 1996; Houde et al. 2009; Cho \& Yoo 2016; Yoon \& Cho 2019).
However, we do not see such a problem with the E/B power asymmetry.
Compare the results for simulation groups {\bf G1} and {\bf G2}, which are denoted by the black and the red curves in Figure \ref{fig:sum}, respectively.
They have similar sonic Mach numbers, but the former has $\sim 2.5$ and the latter has $\sim10$  independent eddies along the LOS.
The fact that simulation groups {\bf G1} and {\bf G2} have similar results verifies that the E/B power asymmetry does not suffer from
an averaging effect arising from independent eddies along the LOS.

The E/B power asymmetry method mainly works for super-Alfv\'enic turbulence.
Then, is super-Alfv\'enic turbulence is a valid model for molecular clouds?
The distribution of $M_A$ in the ISM is not well known.
Nevertheless, there are claims that molecular clouds are globally or locally super-Alfv\'enic (see discussions in Padoan \& Nordlund 1999; Padoan et al. 2010;  Federrath et al. 2016;  Kritsuk, Ustyugov, \& Norman, 2017; Li et al. 2022; Hcar et al. 2022),

In this paper we have demonstrated that the E/B power asymmetry is a steep function of the Alfv\'en Mach number $M_A$
for $ 1\lesssim M_A \lesssim 30$ (see Figure \ref{fig:sum} for solenoidal driving and Figure \ref{fig:comp} for compressive driving; see also fitting formulas for solenoidal driving in Equations (\ref{eq:90}) and  (\ref{eq:30})).
Therefore, we can use the E/B power asymmetry to constrain the strength of the mean magnetic field in supersonic and super-Alfv\'enic turbulence.

\noindent
\\
This work was supported by research fund of Chungnam National University in 2020.





\end{document}